\documentstyle[preprint,tighten,aps,psfig]{revtex}
\begin{document}
\title{The chiral symmetry restoration phase transition
in baryon spectrum}
\author{ L. Ya. Glozman}
\address{ Institute for Theoretical
Physics, University of Graz, Universit\"atsplatz 5, A-8010
Graz, Austria
 \footnote{e-mail: lyg@cleopatra.kfunigraz.ac.at}}
\maketitle

\begin{abstract} 
It is shown that in the large $N_c$ limit the light baryon
spectrum exhibits the chiral restoration phase transition
at high enough excitation energy. Such a phase transition
is evidenced by the systematical parity doublets observed
in the upper part of $N$ and $\Delta$ spectra.
\end{abstract}

\bigskip
\bigskip

At low temperature and density the almost perfect
$SU(2)_L \times SU(2)_R$ global chiral
symmetry of the QCD Lagrangian in the $u,d$ sector
is realized in the hidden Nambu-Goldstone mode. At
higher temperatures or/and nuclear densities the chiral
symmetry restoration occures, the phenomenon which is
well established in the former case on the lattice \cite{Karsch}
(there are little doubts that it also will happen at high
density). Much efforts are being made  to study the
chiral restoration phase transition both theoretically
and experimentally in heavy ion collisions. 
It has been recently speculated \cite{Doublets}
that actually the upper part of  $N$ and $\Delta$ spectra exhibits
the chiral restoration phase transition which is evidenced by
the systematical parity doublets there. The aim of the present note
is to give a general proof that indeed there must happen
the chiral symmetry restoration at some baryon excitation energy.\\

The  $SU(2)_L \times SU(2)_R$ global chiral
symmetry is
equivalent to the independent vector and axial rotations
in the isospin space. The axial transformation mixes
states with different spatial parities. Hence, if this
symmetry of the QCD Lagrangian were intact in the vacuum,
one would observe parity degeneracy of all hadron states
with otherwise the same quantum numbers. This is however
not so and it was a reason for  suggestion in the early days of QCD
 that the chiral symmetry of the QCD Lagrangian is broken down
to the vectorial subgroup $SU(2)_V$ by the QCD vacuum, which 
reflects a conservation of the vector current (baryon number). 
That this is so is directly
evidenced by the nonzero value of the quark condensate

\bigskip
\bigskip

\begin{equation}
<\bar \psi \psi> \simeq -(240 - 250 MeV)^3 
\label{condensate},
\end{equation}

\noindent
which represents the order parameter associated with
the chiral symmetry breaking. The nonzero value of the
quark condensate directly shows that the vacuum state
is not chiral-invariant.\\

Physically the nonzero value of the quark condensate implies
that the energy of the state which contains an admixture
of ``particle-hole'' excitations (real vacuum of QCD) is below
the energy of the vacuum for a free Dirac field, in which
case all the negative-energy levels are filled in and all the
positive-energy levels are free. This can happen only due
to some nonperturbative attractive
Lorentz scalar gluonic interactions between quarks
which pairs the left quarks and the right antiquarks (and vice versa)
in the vacuum.
In perturbation theory to any order the structure of the
trivial Dirac vacuum persists. Such a situation is typical in
many-fermion systems (compare, e.g., with the theory of
superconductivity) and implies that there must appear
quasiparticles with dynamical masses.\\

The physical mechanism for the chiral symmetry restoration at
high temperatures and/or nuclear densities is rather simple.
In the former case it happens because the thermal excitations
of quarks and antiquarks from the vacuum lead to the Pauli
blocking of the levels which are necessary for the formation
of the condensate. In the latter case these levels are
occupied by the valence quarks of the high density quark (nuclear)
matter.\\

For the illustration of these phenomena we will use the Nambu
and Jona-Lasinio model, that adequately reflects the underlying
chiral symmetry of QCD and exhibits the chiral symmetry breaking
(restoration) phase transition \cite{reviews}.\\

Any scalar gluonic interaction between current quarks, which
is responsible for the chiral symmetry breaking in QCD (which
is generally nonlocal), in the
{\it local} approximation is given by the 4-fermion operator
$(\bar \psi \psi)^2$. Because of the underlying chiral invariance
in QCD this interaction should be necessarily accompanied by
the interaction
 $(\bar \psi i\gamma_5 \vec \tau \psi)^2$ with the same strength.
Thus any generic Hamiltonian density in the local approximation is
given by (contains as a part) the NJL interaction model 
(for simplicity we restrict discussion 
 to the u,d flavour sector and to the chiral limit):

\begin{equation}
H= -G\left[(\bar \psi \psi)^2 + 
(\bar \psi i\gamma_5 \vec \tau \psi)^2\right]
\label{NJL}.
\end{equation}

\noindent
If the strength of the effective interaction $G$ exceeds some
critical level
(which happens when the strong coupling constant $\alpha_s(q)$
is big enough)
, then the nonlinear {\it gap} equation

\begin{equation}
M = -2G<\bar \psi \psi> 
\label{gap1},
\end{equation}

\begin{equation}
<\bar \psi \psi> = - \frac{N_c}{\pi^2}
\int_0^{\Lambda_\chi} d|{\bf p}|  {\bf p}^2 \frac{M}{\sqrt({\bf p}^2 + M^2)} 
\label{gap2}
\end{equation}

\noindent
admits a nontrivial solution $ <\bar \psi \psi> \not= 0$, which means
that the initial vacuum becomes rearranged and instead of the
Wigner-Weyl mode of chiral symmetry one obtains the Nambu-Goldstone
one. Thus the appearance of the quark condensate is equivalent
to the appearance of the gap in the spectrum of elementary excitations
in QCD (i.e. of the constituent mass $M$). Hence the constituent quark
is  a quasiparticle in the Bogoliubov sense,
i.e. is a coherent superposition of the bare particle-hole excitations.
 In terms
of the noninteracting constituent quarks the vacuum is again trivial,
but contains a gap 2M. The treatment of the scalar interaction 
between bare quarks in the vacuum in
the Hartree-Fock (mean field) approximation, from which the gap
equation (\ref{gap1})-(\ref{gap2}) is obtained, is equivalent to the
vacuum of {\it noninteracting} constituent quarks. In the vacuum state
the second term of (\ref{NJL}) does
not contribute to the constituent quark self-energy in
the mean field approximation. Contributions beyond the mean field
approximation (i.e. constituent quark self energy due to pion
and sigma loops, which are higher order effects in the $1/N_c$ expansion,
see, e.g., \cite{D})
do not violate significantly the qualitative picture of
the vacuum in the mean field approximation.\\

Consider  the chiral symmetry restoration in the vacuum
at high temperatures. At zero temperature the vacuum condensate
is given by (\ref{gap2}). At finite temperature it is affected
and given by

\begin{equation}
<\bar \psi \psi> = - \frac{N_c}{\pi^2}
\int_0^{\Lambda_\chi} d|{\bf p}|  {\bf p}^2 \frac{M}{E_p} 
[1-n_+(p)-n_-(p)] 
\label{gap3},
\end{equation}

\noindent
where $E_p =\sqrt({\bf p}^2 + M^2)$ and $n_{+-}(p)$ is the 
Fermi-Dirac distribution function for  quarks and antiquarks
at vanishing chemical potential,

\begin{equation}
n_{+-}(p) = \frac{1}{1 + e^{\frac{E_p}{T}}} 
\label{FD}.
\end{equation}

\noindent
In the latter equation $T$ is the temperature. 
At some critical temperature, $T_c$, the nontrivial gap solution
disappears and the chiral symmetry becomes restored. Physically
this is because the thermal excitations of quarks and antiquarks
lead to the Pauli blocking of the levels which are necessary
for the formation of the condensate. The formal (mathematical) reason
is that the quark distribution
function  $n(p)$ is pushed out from the $p=0$ point and becomes broad
and thus affects the gap equation so  that the self-consistent
(gap) solution vanishes.\\

Consider baryons, i.e. the systems that contain valence
quarks on the top of the vacuum. The valence quarks interact
to each other and also to the quarks and antiquarks of the
vacuum. In this case the eq. (\ref{gap3}) should be modified
and the distribution function of valence quarks should be
incorporated (we still assume that the chemical potential
is small so that we can neglect it):

\begin{equation}
<\bar \psi \psi> = - \frac{N_c}{\pi^2}
\int_0^{\Lambda_\chi} d|{\bf p}|  {\bf p}^2 \frac{M}{E_p} 
[1-n_v(p)-n_+(p)-n_-(p)] 
\label{gap4}.
\end{equation}
\\

If the number of colors $N_c$ increases, the valence
quarks in baryons become denser and denser because the size
of baryons is governed by $\Lambda_{QCD}$ which remains fixed.
In this case the Hartree-Fock picture becomes exact and the
system of large $N_c$ valence quarks can be described by the sum
of identical quarks moving in the self-consistent field
of other $N_c - 1$ quarks \cite{Witten}. In the large $N_c$
limit the strong decay of baryons vanishes , which means
that the system of a large amount of valence quarks is in
 thermal equilibrium. This system can obviously be ascribed
some temperature $T$. The valence quarks interact not
only to each other but also to quarks and antiquarks of the
vacuum, which is due to fluctuations of valence quark into
other one plus quark-antiquark pair (Fig. 1).
The diagram of Fig. 1
is not supressed in the large $N_c$ limit and represents an effective
meson exchange between valence quarks in baryons
\cite{GR}\footnote{Note that while these graphs
contribute to baryon mass at the order $N_c$, their
contribution to $N$-$\Delta$ mass splitting appears only
at the order $N_c^{-1}$.}. Because of  intensive interaction
of the valence quarks with the quarks and antiquarks of the
vacuum, the valence quarks 
in baryons are in
 thermal equilibrium with the vacuum (the vacuum plays a
role of a thermostat or vice versa). This means that the same temperature
must be ascribed to both the valence quarks and the vacuum. In particular,
the same temperature persists in all distribution functions
in eq. (\ref{gap4}).\\

The large $N_c$ system of valence quarks is an {\it open}
system. Only together with the quarks and antiquarks (and gluons) 
of the
vacuum it becomes a closed one. Hence the state of the system of valence
quarks is necessarily a {\it mixed} one and the apparatus
of quantum statistics should be used. A measurable physical
baryon with its complete set of quantum numbers, represents a
pure (coherent) state of a closed system (valence quarks plus vacuum).
The mixed state of a system of valence quarks can be obviously
presented as a superposition of all possible pure states
with the same energy.\\

Consider now the ground state baryons. In this case all
valence quarks occupy the same ground state orbital $0s$,
which means that the temperature of both valence quarks and
vacuum is $T=0$. Since  only one level is partially blocked by
the valence quarks, the quark condensate in the present state
is practically the same as the one of the true vacuum (the
true vacuum is the vacuum with no valence quarks on the
top of it).
This is equivalent  to the obvious statement that
the small chemical potential does not perturb much the true
vacuum at zero temperature.\footnote{One should not mix it
with the chiral restoration at high nuclear (quark) densities
at $N_c=3$. In the latter case all the low levels are occupied
by valence quarks.}\\

In the excited baryon there is a coherent superposition 
of one, two, three,... valence quarks that are in the excited states.
Such a pure state of valence quarks plus vacuum
is described by a set of quantum numbers,
in particular by its spin. If one considers a system of valence
quarks only (which is an open one), such a system in the excited
state contains an {\it incoherent} superposition of one, two, three,...
valence quarks that are in the excited states and the temperature
of this system is above zero.
This system is described by the corresponding distribution
function $n_v$.
 Such a system, with the fixed
temperature and energy, can be expanded into the set of baryons
with the same energy and all allowed spins  1/2, 3/2, 5/2,... .
Since the
excited system of valence quarks lives in the thermal equilibrium with its
vacuum, the temperature of the vacuum is also above zero. 
Physically this means that the true vacuum becomes strongly perturbed
by the excitation of the baryon. At some critical excitation
energy of the baryon (i.e. at some critical temperature)  
its vacuum undergoes the
chiral restoration phase transition. In the example considered
above it follows from the fact that at some temperature the
self-consistent nontrivial solution of eqs. 
(\ref{gap1}) and (\ref{gap4}) vanishes. Around this excitation
energy (temperature) there must appear systematical approximate
parity doublets  with all possible spins.\\

The argument above is robust and general and does not rely
on NJL model, which is used only for illustration. The only
important element is that when the number of colors increases,
the valence quarks in baryons can be ascribed a temperature and
this system lives in thermal equilibrium with its vacuum.\\

How close to reality is the large $N_c$ picture of baryons? It
has been shown \cite{MDJ} that in the large $N_c$ limit the
$SU(6)$ symmetry of baryons becomes exact (to be precise, 
the low-energy observables such as baryon masses, magnetic moments
and axial coupling constants are described by the corresponding
$SU(6)$ symmetrical wave functions). As it is well known the $SU(6)$
symmetry
works reasonably well for all these observables. For instance,
the $SU(6)$ predictions for baryon magnetic moments are
satisfied at the level 15-20\%. The $N$-$\Delta$ splitting is
of the order 25-30\% of the baryon mass. This implies that
one can expect predictions of the large $N_c$ limit 
to be correct at the
level 20-30\%. This confidence level is enough to anticipate that
the  arguments above for the chiral symmetry restoration
at large $N_c$ will survive in the real world with $N_c=3$.\\

How about mesons? In the latter case even in the large $N_c$
limit the meson still consists of one valence quark and antiquark,
so the thermodynamical description
 cannot be used here for valence degrees of freedom. In addition
in the present case there are no diagrams in the large $N_c$
limit, similar to that one of Fig. 1. This implies that there
is no mutual impact of the valence degrees of freedom and of the
quarks and antiquarks of the vacuum. Hence the meson 
spectra should not show the chiral symmetry restoration and 
parity doublets.\\

The still poorly mapped upper part of the light baryon spectrum
\cite{Caso} exhibits  remarkable parity doublet patterns. To these belong
$N(2220), \frac{9}{2}^+ - N(2250), \frac{9}{2}^-$, 
$N(1990), \frac{7}{2}^+ - N(2190), \frac{7}{2}^-$,
$N(2000), \frac{5}{2}^+ - N(2200), \frac{5}{2}^-$,
$N(1900), \frac{3}{2}^+ - N(2080), \frac{3}{2}^-$,
$N(2100), \frac{1}{2}^+ - N(2090), \frac{1}{2}^-$,
$\Delta(2300), \frac{9}{2}^+ - \Delta(2400), \frac{9}{2}^-$,
$\Delta(1950), \frac{7}{2}^+ - \Delta(2200), \frac{7}{2}^-$,
$\Delta(1905), \frac{5}{2}^+ - \Delta(1930), \frac{5}{2}^-$,
$\Delta(1920), \frac{3}{2}^+ - \Delta(1940), \frac{3}{2}^-$,
$\Delta(1910), \frac{1}{2}^+ - \Delta(1900), \frac{1}{2}^-$.
 The splittings within the parity partners are typically
within the 5\% of the baryon mass. This value should be given
a large uncertainty range because of the experimental uncertainties
for the baryon masses of the order 100 MeV. Only a couple of states
in this part of the spectrum do not have their parity partners so
far observed. The low energy part of the spectrum, on the
other hand, does not show this property of the parity doubling.
The increasing amount of the near parity doublets in the high
energy sector is an evidence of the chiral symmetry resoration.\\

If chiral and deconfinement phase
transitions coinside, the conclusion should be that the
highly excited baryons with masses above some critical value
(where the phase transition is completed) should not exist
because deconfinement phase transition should be dual to a
very extensive string breaking at big separations of colour
sources (colour screening). 
Whether this point corresponds to approximately 2.5 GeV or
higher should be answered by future experiments on high
baryon excitations. The phase transition can be rather
broad because of the explicit chiral symmetry breaking
by the nonzero value of current quark masses.\\

There is a couple of the well confirmed  states
 $N(2600), \frac{11}{2}^-$ and 
$\Delta(2420), \frac{11}{2}^-$, in which case the parity partners 
are absent \cite{Caso}. Thus it will be rather important to try to find
them experimentally.\\

{\bf Figure captions}

Fig.1 An effective meson exchange diagram which represents a
thermodynamical exchange between valence quarks in baryons and
the quarks and antiquarks of the vacuum.

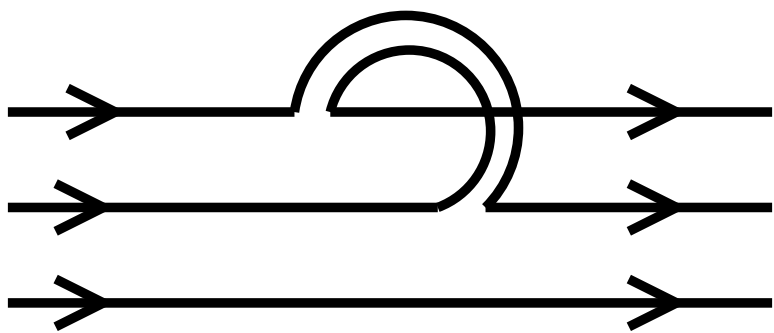
\end{document}